\documentclass[pre,twocolumn,aps]{revtex4}

\usepackage{graphicx}
\usepackage{amsmath}
\usepackage{amsfonts}
\usepackage{amssymb}
\usepackage{epsf}

\begin{document}

\title{
From subdiffusion to superdifffusion of particles on solid surfaces
}
\author{
A. M. Lacasta$^{1}$, J. M. Sancho$^{2}$,
A. H. Romero$^{3}$, I. M. Sokolov$^{4}$, and
K. Lindenberg$^{5}$
}
\affiliation{
$^{(1)}$
Departament de F\'{\i}sica Aplicada,
Universitat Polit\`{e}cnica de Catalunya,
Avinguda Doctor Mara\~{n}on 44, E-08028 Barcelona, Spain\\
$^{(2)}$
Departament d'Estructura i Constituents de la Mat\`eria,
Facultat de F\'{\i}sica, Universitat de Barcelona,
Diagonal 647, E-08028 Barcelona, Spain\\
$^{(3)}$
Advanced Materials Department, IPICyT, Apartado
Postal 3-74 Tangamanga, 78231 San Luis Potos\'\i , SLP, M\'exico\\
$^{(4)}$ 
Institut f\"{u}r Physik, Humboldt Universit\"{a}t zu Berlin,
Newtonstr. 15, 12489 Berlin, Germany\\
$^{(5)}$
Department of Chemistry and Biochemistry 0340, and Institute for
Nonlinear Science,
University of California, San Diego, La Jolla, California 92093-0340, USA\\
}

\date{\today}

\begin{abstract}

We present a numerical and partially analytical study of classical
particles obeying a
Langevin equation that describes diffusion on a
surface modeled by a two dimensional potential.  The potential
may be either periodic or random.  Depending on the potential and the
damping, we observe superdiffusion, large-step diffusion, diffusion,
and subdiffusion.  Superdiffusive behavior is associated with 
low damping and is in most cases transient, albeit often long.
Subdiffusive behavior is associated with highly damped particles
in random potentials.  In some cases subdiffusive behavior persists
over our entire simulation and may be characterized as metastable. 
In any case, we stress that this rich variety of behaviors emerges
naturally from an ordinary Langevin equation for a system described
by ordinary canonical Maxwell-Boltzmann statistics.
\end{abstract}

\maketitle

PACS: 0.54-a, 68.43.Jk, 68.35.Fx

\section{Introduction}

Diffusion processes of atoms, molecules, and clusters of
molecules on surfaces have been subjects of research for many decades
due to their intrinsic interest and their technological importance.
Some examples of modern applications include self-assembled molecular
film growth, catalysis, and surface-bound nanostructures~\cite{nano,ala}. 
Also, many techniques that are used to
characterize the growth of surfaces are based on the diffusion and subsequent
adsorption of particles or molecules. These applications have led to a recent
resurgence of interest in such processes, but now involving the motion of
small and large organic molecules~\cite{phd,prl1} and of adsorbed metal
clusters composed of tens and even hundreds of atoms~\cite{phystoday,prl2}.

Recent research activity has been both experimental and
theoretical~\cite{nano,ala,phd,prl1,phystoday,prl2,jensen,levi,older,henry,condmat,chen},
the principal questions in these studies being the determination of the
jump lengths of large molecules or clusters on surfaces, and the
diffusion asymmetry
along some surface crystallographic directions or along particular paths
when there are many obstacles such as steps and/or impurities to negotiate.
One of the more exciting experiments has involved the direct observation
of the surface mobility of fairly large organic molecules using a newly
devised method of independently measuring the mean square displacement
and the hopping rate of these molecules, and then taking advantage of a
well known relation between them to extract the root mean square jump
lengths~\cite{prl1}.  These experiments seem to indicate that, as with
single atoms in some cases~\cite{oneatom,ferron}, long jumps spanning
multiple lattice spacings may play a dominant role in the diffusion of these
molecules.  Another ``experiment'', this one numerical (molecular
dynamics),  has led to the  prediction that clusters of hundreds of Au atoms
will exhibit L\'{e}vy-type power-law flight length and sticking time
distributions~\cite{prl2}.  
On the other hand, the theoretical literature on the subject of
surface diffusion tends to
be rather complex and, most importantly, tends to invoke
L\'{e}vy  walks or flights~\cite{mike} as a model {\em input}.
While these approaches can
provide insights on the effects of other elements of the model on
the dynamics of particles on surfaces, they provide little insight on
how the motion involving long jumps comes about in the first place.  
These simulations, which aim to reproduce realistic 
systems, do not stress the generic, minimal ingredients
of such behavior.

While a microscopically accurate analysis of surface diffusion requires
extensive calculations (e.g., {\em ab initio}, or molecular dynamics),
even the most powerful currently available computers
can not carry such calculations to anywhere near experimentally relevant
time scales~\cite{ferron,gross}.  Moreover, current experimental
probes of the topography of surfaces, scanning tunneling
microscopy and atomic force microscopy,
are usually carried out at relatively high temperatures, which leads to
additional difficulties for first-principles calculations.
Therefore, simpler approaches are essential and valuable~\cite{theory}.
The scenario developed in this paper is of the simple Langevin type.  It
models classical particles moving in a two-dimensional potential, periodic or
random, under the action of $\delta$-correlated Gaussian thermal
fluctuations and the associated linear dissipation, the important
control parameter of the model being the friction coefficient. 
In our search for different behaviors as a function of the control
parameter, we must be mindful of the fact that experimental
observations
are not necessarily strictly {\em asymptotic}, that is, that a
particular
behavior may be transient.  Transient behaviors may of course
persist for very long times, but to find them theoretically it is not
enough to carry out an asymptotic analysis. We find that
in spite of the simplicity of the
model, it is able to reproduce the entire range of
experimentally and computationally observed phenomenologies, ranging
from superdiffusion through large-step diffusion all the way to
subdiffusion.  In particular, we show that L\'{e}vy-like statistics appear
quite naturally within the usual Langevin scheme for underdamped motion
in a periodic potential.  It is important to 
stress that in this case the L\'{e}vy-behavior emerges as an
\emph{intermediate} asymptotic for an ensemble of particles with a
Maxwell-Boltzmann distribution of velocities.

Our paper is organized as follows.  In Sec.~\ref{sec:model} we present
the Langevin model and briefly list the quantities to be considered in
subsequent sections to characterize the motion of an ensemble of
particles on a surface. In Sec.~\ref{sec:numericalperiodic} numerical
results are presented for the periodic surface potential, as are
analytic results for the dependence of the diffusion coefficient on
friction obtained in detail in the Appendix. 
Section~\ref{sec:numericalrandom}
shows our results for the random surface potentials. Finally, we
conclude with a summary and some future directions in
Sec.~\ref{sec:conclusions}.

\section{The model}
\label{sec:model}

We study the diffusion of a particle in a two-dimensional
potential $V(x,y)$ of characteristic length scale $\lambda$,
in the presence of thermal noise and the associated
dissipation.  The model is embodied in the equations of motion,
\begin{equation}
\begin{split}
m\ddot{x} &= -\frac{\partial}{\partial x}
V\left(\frac{x}{\lambda},\frac{y}{\lambda}\right)
-\mu\dot{x}+\xi_x(t)\\
m\ddot{y} &= -\frac{\partial}{\partial y}
V\left(\frac{x}{\lambda},\frac{y}{\lambda}\right)
-\mu\dot{y}+\xi_y(t),
\end{split}
\label{eq:1}
\end{equation}
where $m$ is the mass of the particle and a dot denotes a derivative
with respect to $t$. The generalization to
distinct length scales $\lambda_x$ and $\lambda_y$ is straightforward.
The parameter $\mu$ is the coefficient of friction, and the
$\xi_i(t)$ are mutually uncorrelated white noises that obey the
fluctuation-dissipation relation,
\begin{equation}
\langle \xi_i(t)\xi_j(t')\rangle =2\mu k_BT\delta_{ij}\delta(t-t').
\end{equation}

Equations~(\ref{eq:1}) can be written in terms of the scaled dimensionless
variables
\begin{equation}
r_x=\frac{x}{\lambda}, \qquad r_y=\frac{y}{\lambda},\qquad
\tau=\sqrt{\frac{V_0}{m}}\frac{t}{\lambda},
\end{equation}
where $V_0$ is some measure (e.g. the maximum or the mean
maximum) of the potential.
In terms of these variables, and with a dot now denoting a derivative
with respect to $\tau$,
\begin{equation}
\begin{split}
\ddot{r}_x &= -\frac{\partial}{\partial r_x} {\cal {V}}(r_x,r_y)
-\gamma\dot{r}_x+\zeta_x(\tau)\\
\ddot{r}_y &= -\frac{\partial}{\partial r_y} {\cal {V}}(r_x,r_y)
-\gamma\dot{r}_y+\zeta_y(\tau),
\end{split}
\label{eq:2}
\end{equation}
where ${\cal{V}}(r_x,r_y)= V(x/\lambda,y/\lambda)/V_0$ is
the dimensionless potential, and the scaled
noise obeys the fluctuation-dissipation relation,
\begin{equation}
\langle \zeta_i(\tau)\zeta_j(\tau')\rangle =2\gamma {\mathcal T}
\delta_{ij}\delta(\tau-\tau').
\end{equation}
This scaling serves to stress that there are only two independent
parameters in this model, the scaled temperature ${\cal T}$ and the
scaled dissipation $\gamma$: 
\begin{equation}
{\mathcal T}= k_BT/V_0, \qquad
\gamma=\mu\lambda/\sqrt{mV_0}.
\end{equation}
The scaled temperature will be fixed
at a value smaller than unity so that the thermal fluctuations do not
overwhelm the potential, but not too much smaller than unity so that
activated passage over potential barriers is possible within a
reasonable time.  

The model~(\ref{eq:1}) is a ``standard'' Brownian motion model.
It relies on well-established ideas of statistical mechanics and invokes
nothing special about the fluctuations and the dissipation. The random
forces are normal thermal fluctuations, Gaussian and
$\delta$-correlated, and the dissipative forces that accompany the
fluctuations are constructed so as to insure 
thermal equilibration.  Nevertheless, we argue that many of the
dynamical features of a particle evolving under these equations of
motion have not been investigated until recently~\cite{ourletter}.
Furthermore, we assert that it is not necessary
to inject special assumptions such as L\'{e}vy flights or special memory
effects into models of surface diffusion, but that these features appear
naturally from this standard model, thus placing
the entire range of phenomena observed in surface diffusion on an
equal common footing. Furthermore, since
recent experimental and theoretical results have been presented
for a variety of surfaces, we explore periodic as well as
random surface potentials, the latter generated
according to a given distribution and with a given short-range
spatial correlation.  

The most straightforward quantity to characterize the surface diffusion 
process is the familiar mean square displacement of the particle,
\begin{equation}
\langle r^2(\tau)\rangle = \langle [r_x(\tau)-r_x(0)]^2\rangle + \langle
[r_y(\tau)-r_y(0)]^2\rangle.  
\end{equation}
Normal diffusive behavior is characterized by a linear dependence
on time, $\langle r^2(\tau)\rangle \sim D\tau$, where $D$ is the
diffusion coefficient, while
non-diffusive behavior shows a different time dependence, $\langle
r^2(\tau)\rangle \sim \tau^\alpha$, with $\alpha<1$ for subdiffusive behavior
and $\alpha>1$ for superdiffusive motion.
Not only do we seek to establish the nature of the motion as a
function of the friction coefficient as measured by moments such as the
mean square displacement, but we seek to establish a theoretical
framework that will allow us to fully understand and predict the nature
of the motion under realistic circumstances that may include more
complex geometries.  Toward this purpose we will explore the dependence
of the diffusion coefficient $D$ on friction at large time scales.

Another characterization of the process at intermediate time
scales is obtained from the probability density function (pdf)
$P(r,\tau)$ of particle displacements $r$ at time $\tau$.  In
particular, this pdf reflects the long stretches of ballistic motion
evident in the intermediate time dynamics in the low friction regime.
Indeed, long ballistic excursions lead to a specific behavior of this
distribution that in turn implies that the velocity
of these particles remains correlated over
considerable time intervals.  The best way to
characterize such correlations is in terms of the velocity power spectrum
$S(\omega )=\langle \mathbf{v}(\omega )\mathbf{v}(-\omega )\rangle$.
We note that this spectrum is directly connected with the asymptotic
diffusion coefficient of the particle. According to the Taylor-Kubo formula,
$D=\int_{0}^{\infty }C(\tau)d\tau$,
where $C(\tau)=\langle \mathbf{v}(\tau^{\prime
})\mathbf{v}(\tau^{\prime }+\tau)\rangle $ is the velocity-velocity
correlation function. The latter is connected with $S(\omega )$ by the
Wiener-Khinchin relation, $S(\omega )=\int C(\tau)e^{i\omega \tau}d\tau$.
These various observables are studied below for both periodic and random
surface potentials.

\section{Periodic potential}
\label{sec:numericalperiodic}
 
\begin{figure}
\begin{center}
\includegraphics[width = 9cm]{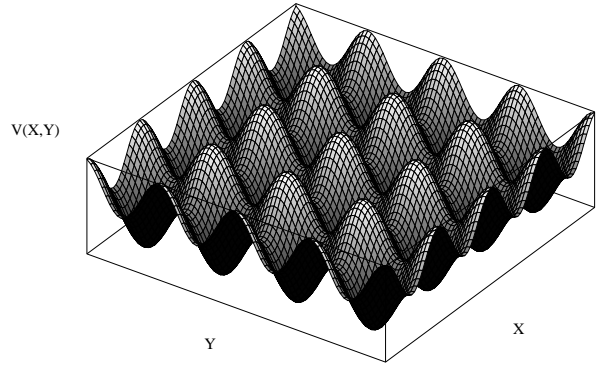}
\end{center}
\caption{
A finite portion of the much larger two-dimensional periodic potential
in which a particle diffuses.
}
\label{fig1}
\end{figure}

\begin{figure}
\begin{center}
\includegraphics[width = 8cm]{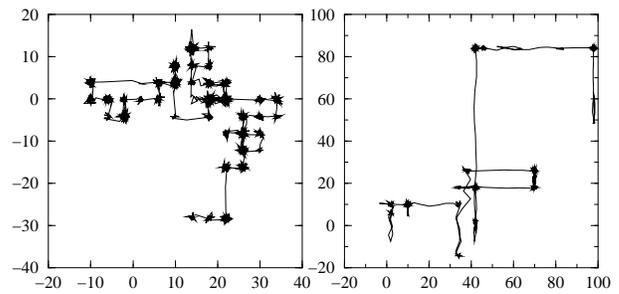}
\end{center}
\caption{
Left: A trajectory for $\gamma=1$ over $t=20,000$ time units.
Right: A trajectory for $\gamma=0.04$ over $t=15,000$ time
units. The period of the potential is $\lambda=4$. Note the different
scales in the two panels.
}
\label{fig2}
\end{figure}

We consider the periodic potential
\begin{equation}
V(x,y)=V_0 \cos\left(\frac{\pi x}{\lambda} + \frac{\pi y}{\lambda}\right)
\;\cos\left(\frac{\pi x}{\lambda} - \frac{\pi y}{\lambda}\right),
\end{equation}
which has maxima at positions $(n\lambda,m\lambda)$ and minima at
$((n+\frac{1}{2})\lambda,(m+\frac{1}{2})\lambda)$, where $n$ and $m$ are
integers.  The barrier height at the saddle points is $V_0$.  This
potential is shown in Fig.~\ref{fig1}.
In Fig.~\ref{fig2} we show examples of trajectories obtained for
two different friction coefficients upon numerical simulation
of the equations of motion. 
In our simulation we have used $\lambda=4$ and $m=1$, with a fixed scaled temperature 
${\cal T}=0.2$ and the scaled dissipation as the independent parameter.
Note that since the potential can be generated
analytically and the equations of motion are continuous, the system is
infinite and it is not necessary to specify boundary conditions. 
One of the trajectories [panel (a)] is for large friction, and the
particle is seen to follow what appears to be typical diffusive motion
characterized by short steps of length $\lambda$ along the
crystallographic directions.  The other trajectory
[panel (b)] corresponds to small friction and clearly shows the
preponderance of long ($\gg \lambda$) tracks along one
crystallographic direction before turning to another.  

\begin{figure}
\begin{center}
\includegraphics[width = 8.5cm]{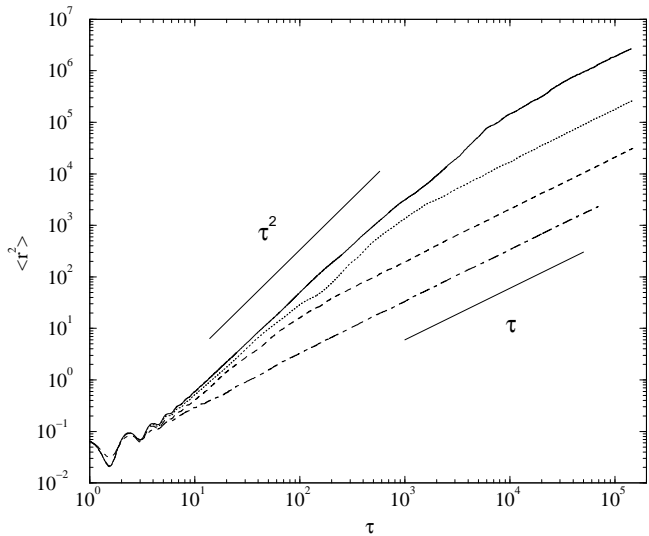}
\end{center}
\caption{
Mean square displacement vs time for an ensemble of 5000 particles in the
periodic potential with a Maxwell-Boltzmann initial distribution. 
$\gamma=0.0004$ (solid line), $0.004$ (dotted line),
$0.04$ (dashed line) and $0.4$ (dot-dashed line).
}
\label{fig3}
\end{figure}

\begin{figure}
\begin{center}
\includegraphics[width = 8.5cm]{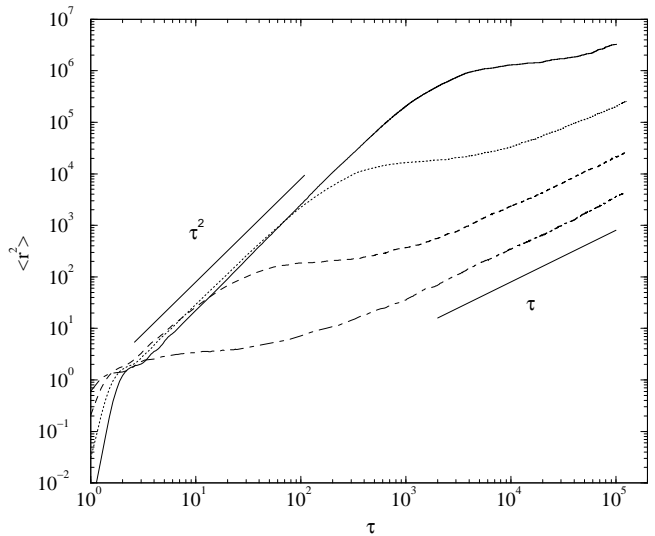}
\end{center}
\caption{
Mean square displacement vs time for an ensemble of 5000 particles in the
periodic potential with zero velocity, and located at the origin.
$\gamma=0.0004$ (solid line), $0.004$ (dotted line),
$0.04$ (dashed line) and $0.4$ (dot-dashed line).
}
\label{fig4}
\end{figure}

The evolution of the mean square displacement
$\langle r^2 \rangle$, averaged over a set of $5000$ particles,
is shown in Figs.~\ref{fig3} and ~\ref{fig4} for several friction
coefficients and different initial conditions.  In Fig.~\ref{fig3}
the particles are initially deposited in a square of side $2\pi$ around
the center of the system according to a Boltzmann equilibrium
distribution for the positions and for the velocities, while in
Fig.~\ref{fig4} they are initially located at the center of the system with 
zero initial velocity. 
Differences in the two cases are observed at short times.
At very long times the memory of
the initial condition is lost, and in all cases the 
motion is diffusive, as expected. Regardless of initial condition,
for small $\gamma$ and at intermediate times there is clear
\emph{superdiffusive} ballistic ($\alpha=2$) behavior
over several decades of time, reflective of the long straight
stretches seen in the low-$\gamma$ trajectory in
Fig.~\ref{fig1}. The ballistic behavior is observed over a time range of
$O(\gamma^{-1})$, and one might be tempted to conclude that it is
therefore a trivial generalization of the motion of a free damped
particle.  However, the situation here is quite different. The
motion is of an ensemble of particles whose initial energy distribution
is either the Boltzmann distribution appropriate to the temperature
${\cal T}$ (Fig.~\ref{fig3}), or all of which initially have zero velocity
(Fig.~\ref{fig4}),
and the mean square displacement is an average over all the particles.
The particles with energies lower than the barrier
height, which includes most of the particles, are at first trapped in
the original potential well.  Others (those with a higher initial
energy, and those that get kicked up sufficiently in energy by the
thermal fluctuations) move out of one well only to be quickly
trapped again, perhaps into a neighboring well.  A few particles, those
that either start with or acquire the highest velocities according to the
Maxwell-Boltzmann distribution, escape the initial well and move over 
longer distances even as they slow down through friction, until they are once
again trapped.  That this complex ensemble behavior should give rise to
a mean square displacement that behaves as shown in Fig.~\ref{fig3}
is not self-evident, and in fact can only be understood in more detail
if we analyze the distribution of particle displacements as a function
of time.  We do so later in this section.

\begin{figure}
\begin{center}
\includegraphics[width = 8cm]{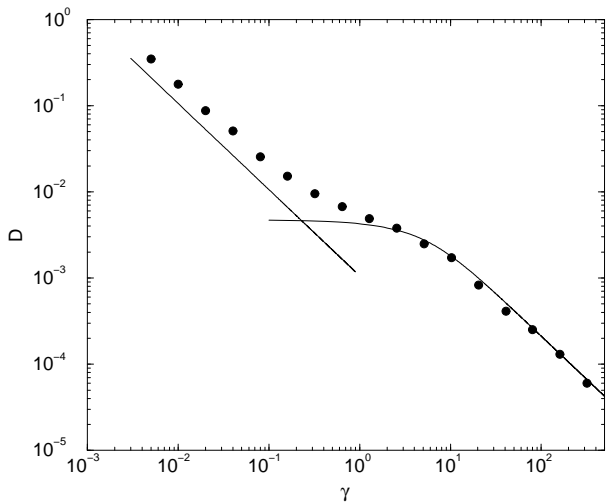}
\end{center}
\caption{
Diffusion coefficient as a function of $\gamma$ for the periodic potential. 
}
\label{fig5}
\end{figure}

As noted above, for very long times the motion is diffusive in all cases.
The diffusion coefficient $D$, evaluated numerically at long time as 
\begin{equation}
D=\lim_{\tau\rightarrow \infty}\frac{<r^2(\tau)>}
{4 \tau},
\label{eq4}
\end{equation}
is shown in Fig.~\ref{fig5} as a 
function of the friction $\gamma$.
The solid lines correspond to theoretical predictions evaluated
in the Appendix.  There we show that
in the overdamped regime [large $\gamma$, see Eq.~(\ref{diffusionover})]
\begin{equation}
D \sim \frac{\pi}{\gamma} \exp \left(-\frac{1}{{\cal T}}\right).
\end{equation}
In the opposite limit [small $\gamma$, see Eq.~(\ref{diffusioninfra})], 
we have
\begin{equation}
D \sim \frac{\pi {\cal T}}{4\gamma}\exp \left(-\frac{1}{\cal T}\right).
\end{equation}
Both predictions fit our numerical results rather well.  Note that there
are \emph{no free parameters} in these results.  Note also that while $D\sim
\gamma^{-1}$ at both high and low friction, the physical reasons are
quite different.  As seen in the Appendix, at high friction the mean
square distance traveled between trapping events is unity independently
of $\gamma$, and the mean time to escape a well and be trapped in
another is proportional to $\gamma$. 
At low friction the mean square distance is proportional to
$\gamma^{-2}$ and the mean time goes as $\gamma^{-1}$.  The ratio of the two
is therefore in any case proportional to $\gamma^{-1}$.  We note that
the low-$\gamma$ result is not applicable to an arbitrary periodic
potential~\cite{chen}.  In particular, it is correct if, as in
our potential, the
direction in which the particle crosses the saddle point lies along the
direction of steepest descent.  If these do not coincide, the particle
can not simply move along effectively one-dimensional channels, and the
low-$\gamma$ behavior of the diffusion coefficient may be modified.

\begin{figure}
\begin{center}
\includegraphics[width = 9cm]{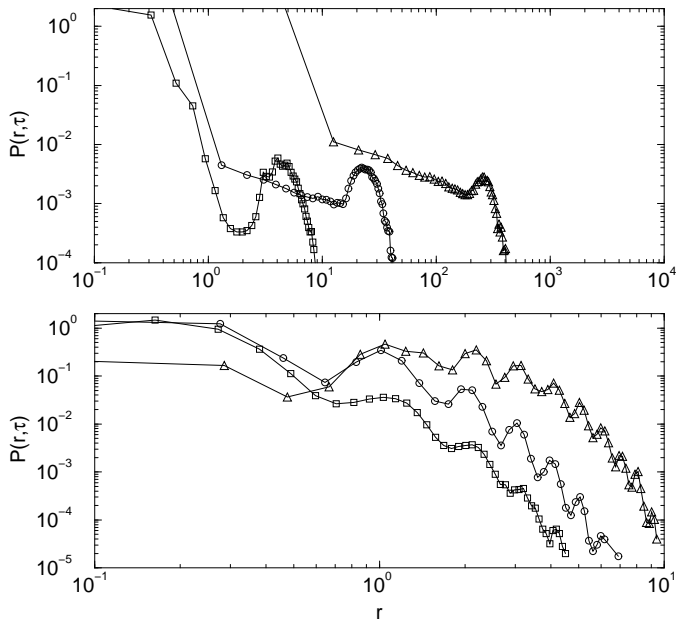}
\end{center}
\caption{
Log-log plot of $P(r,\tau)$ for $\gamma=0.0004$ (top) and for $\gamma=1$ (bottom) and 
three
different values of
time intervals : $\tau= 20$ (squares),
$100$ (open circles), $1000$ (triangles). Note the strong difference in scales.
}
\label{fig6}
\end{figure}

The diffusion coefficient characterizes only the very long-time
asymptotic dynamics.  To characterize the process at intermediate time
scales we introduce the probability distribution function (pdf)
$P(r,\tau)$ of particle displacements $r$ at time $\tau$.  
In particular, this pdf reflects the long stretches of ballistic motion
evident in the intermediate time dynamics in the low friction regime.
The pdf is shown in Fig.~\ref{fig6} (top) for $\gamma=0.0004$ and
three different time intervals $\tau$ for an initially equilibrated
ensemble.  For comparison, we
also show typical pdf's for high damping ($\gamma=1$, bottom) for the
three time intervals.
In the high-$\gamma$ curves the highest maximum
corresponds to no jumps (by far the most likely event at short times).
The next is associated with jumps to a
nearest neighbor well, and so on.  In contrast,
the low-$\gamma$ curves show a
very different behavior, with features
strongly resembling those of a L\'evy-walk
model~\cite{Klafterb,Klaftera}:
a peak at small displacements, a power-law intermediate regime,
and a side hump at high displacements.
Each of these is a
distinct signature of L\'evy-walk-like dynamics, but one must be
cautious in the detailed interpretation of these components, because
in our results they are not attributable to exactly the same sources as
in the L\'evy walk, nor do the exponents fit the L\'evy walk scheme
directly. 
In our system the persistent small displacement peak 
is associated with long trapping periods during which
a particle does not move at all because its energy is not sufficient
to overcome the barrier.  The high displacement peak, which
moves outward with velocity of order unity, is associated
with ballistic motion of those particles that
acquire enough energy to move (and lose it very slowly).
Genuine L\'evy-walk dynamics also exhibit a low displacement peak
and a superdiffusive peak. In the L\'evy walk, the persistent small 
displacement peak stems from the distribution of excursion lengths, whereas the
high displacement peak is associated with laminar events in which a
particle started moving before the observations started and are still
moving without interruption at time $t$.

These small- and large-displacement
domains are separated by a power law behavior. However, here
some important differences must also be stressed.
In the L\'evy walk there is a particular relation between the exponent
$\alpha$ of the time in the mean square displacement $\langle
r^2\rangle \sim t^\alpha$ and the slope $\mu$ of
the power law regime of the distribution, $P(r,\tau)\sim \tau/r^\mu$,
namely, $\alpha=4-\mu$.  Furthermore, the distribution with the features
that we are describing is observed in the regime $2<\mu<3$, that is,
$1<\alpha<2$.  In our distribution the intermediate power law regime
reflects ballistic transport, and yet  the slope in our power law regime
is smaller than unity (approximately $0.7$). 
Nevertheless, the qualitative features
of our distribution track those of the L\'evy walk. 
Our side hump is strongly broadened whereas the
side hump in the L\'evy-walk model is associated with motion at a
single constant velocity.  In our case the
velocity varies according to the equilibrium
Maxwell-Boltzmann distribution.
Note that
the existence of the pronounced side
hump moving with the velocity of the order of unity
reflects the fact that the particles performing long steps
(``flights'') are those in the tail of the Maxwellian velocity distribution. 
This contribution to the pdf in a potential system
arises from a small subset of particles and is thus to be
distinguished from that of a typical underdamped free Brownian
particle~\cite{chaikin}.  
In summary, in our underdamped system the
slower particles are trapped and localized around a potential minimum.  
They do not contribute to transport and lead to
the first peak in the pdf $P(r,\tau)$.  Particles with higher energies 
begin moving essentially ballistically, thus contributing to the side
hump of the pdf that moves toward the right with increasing times in
the first panel in Fig.~\ref{fig6}.  These particles eventually become
trapped due to friction, and these trapping events lead to the power law
portion of the pdf.  Fewer and fewer particles (only those in the
ever smaller highest-velocity tails of the distribution) continue moving
ballistically. The side hump thus becomes narrower and lower with
time.  At long times, $\tau\gg 1/\gamma$, both the long trapping and
ballistic features are of course no longer present as diffusive motion
dominates the behavior.

\begin{figure}
\begin{center}
\includegraphics[width = 9cm]{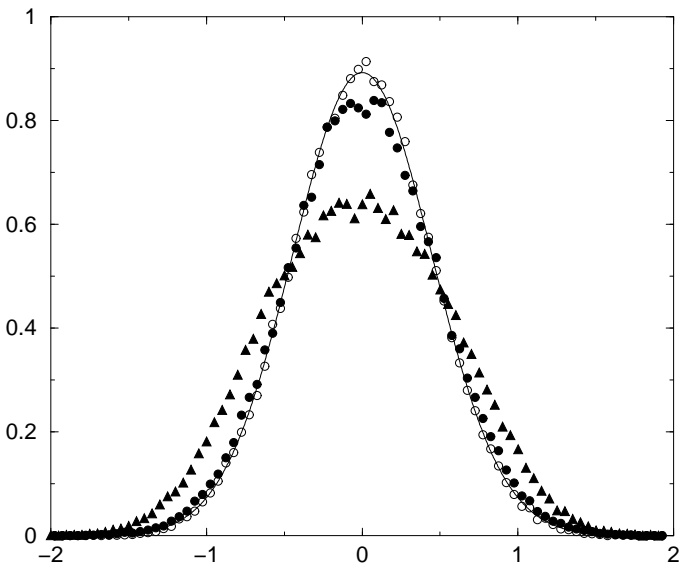}
\end{center}
\caption{
Probability distribution of the velocity for two different
initial conditions compared with a Maxwell distribution (solid curve)
for the same temperature. Triangles: distribution at $\tau=50$ for
zero velocity initial condition.  Circles: distributions at
$\tau=100$ for zero velocity (solid circles) and Maxwell (open
circles)
initial distributions ($\gamma=0.04$).
}
\label{fig7}
\end{figure}

We emphasize that the behavior described above is universal and is
not strongly dependent on whether the initial conditions are equilibrium
or zero-velocity ones (although the exact heights of the peaks do). 
In Fig.~\ref{fig7} the velocity probability distributions of our system at
finite times are compared to a Maxwell distribution. When we implement
Maxwell-Boltzmann initial conditions, the overall velocity 
distribution obtained in our simulations
by time-sampling is purely Maxwellian at all times. For the
nonequilibrium initial conditions where we start with zero velocity at
the saddle point of the potential, some nonequilibrium
transient phenomena are
observed in the small deviations from the Maxwell form. 
We also note that the velocity distribution thermalizes
at times of the order of $10\gamma^{-1}$.
The times in Fig.~\ref{fig7} correspond to $t=2\gamma^{-1}$ and
$4\gamma^{-1}$.

\begin{figure}
\begin{center}
\includegraphics[width = 9cm]{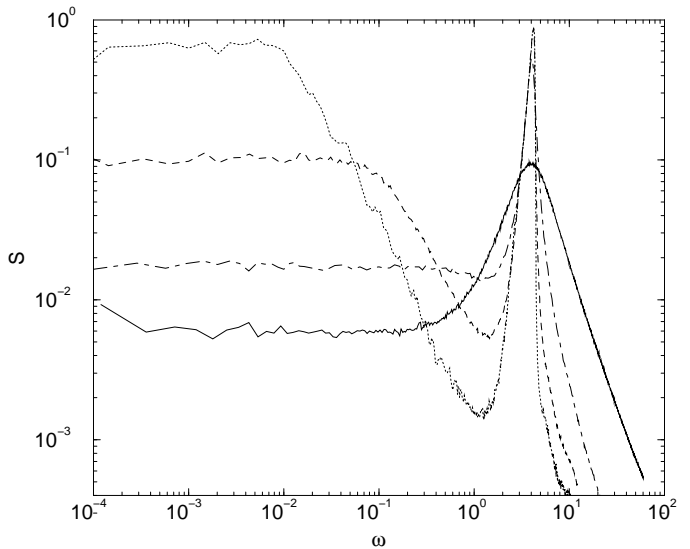}
\end{center}
\caption{$S(\omega)$ for $\gamma=0.004$, $0.04$, $0.4$ and
$4$ in decreasing order.
}
\label{fig8}
\end{figure}

The long, ballistic excursions that lead to the power law and side hump
contributions to the particle displacement distribution function
imply that the velocities of these particles remain correlated over
considerable time intervals.  This aspect is most directly  
characterized by a study of the velocity power spectrum
$S(\omega )=\langle \mathbf{v}(\omega )\mathbf{v}(-\omega )\rangle$.
In the case of diffusive motion one can extract the diffusion coefficients as,
\begin{equation}
D=\frac{1}{4}\left( S_{xx}(\omega=0) + S_{yy}(\omega=0) \right),
\end{equation}
where $S_{ii}(\omega)$ is the power spectrum of the cartesian
component $i$ of the velocity vector.
The behavior of $S(\omega )$ for different
values of the friction coefficient $\gamma $ is shown in Fig.~\ref{fig8}.
The structure of the power spectrum mirrors the 
underlying dynamics. Thus, an evident feature of $S(\omega)$ is
the existence of pronounced
peaks at the frequency $\omega_0 =\pi \sqrt{2}$ of small
oscillations in one well performed by particles trapped in a well.  At
low frequencies, $\omega_0 \ll 1$, a power-law growth of $S(\omega)$ is
observed. This is reflective of the persistent time correlations
associated with ballistic excursions. The slope $\sim 1$ of this
growth corresponds to $C(\tau)\approx const$ and to the ballistic
growth of the mean square displacement. At even smaller frequencies
the power spectrum crosses over to
$S(\omega)=const=4 D$, indicating full decorrelation and emergence of
pure diffusion.

\section{Random potential}
\label{sec:numericalrandom}

\begin{figure}
\begin{center}
\includegraphics[width = 9cm]{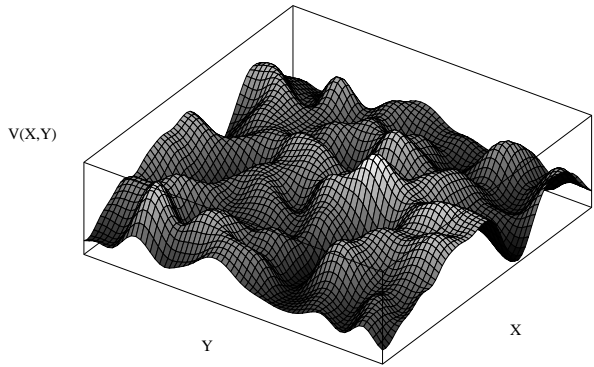}
\end{center}
\caption{
Random potential with the same average potential height as in the
periodic case.
}
\label{fig9}
\end{figure}

Surfaces are usually not completely crystalline or regular because of
the presence of vacancies, defects, and other types of disorder.
One type of surface disorder is represented by a
random potential with spatial correlations that can model
the presence of some spatial finite-ranged order.
An algorithm that can be used to generate surfaces with any given
spatial correlation has been presented in
Refs.~\cite{theirvarious,ourvarious,romero2}.
We have implemented this algorithm for a random potential surface with a
Gaussian distribution and exponential correlation function.  In terms of
the vectors ${\bf x} =(x,y)$ and ${\bf x'} = (x',y')$, our
potential surface has the correlation property,
\begin{equation}
\langle V({\bf x}) V ({\bf x'})\rangle = g({\bf x}-{\bf x'}),
\end{equation}
with
\begin{equation}
g({\bf x}-{\bf x'})=\frac{\varepsilon}{2\pi \lambda'^2} e^{-|{\bf x}-{\bf
x'}|^2/2\lambda'^2},
\end{equation}
which is parameterized by the intensity $\varepsilon$ and the
characteristic length $\lambda'$ that we take to be the same as the
length scale in the periodic potential ($\lambda'=4$).  Note that
the average potential
height is a combination of the parameters.  A typical surface generated
with this algorithm and of average potential height $V_0$ equal to that
of the periodic potential is shown in Fig.~\ref{fig9}. 
In our particular case, we have done our simulations on a square
grid of size  $L=4096$ with $\delta x = 1.0$.
Even though our potential is generated
on a grid, extrapolation of forces can be used to calculate the force
at any point, which is a necessary step in the dynamical
simulation.  We have used the parameter $\varepsilon=100$.
In this case we must generate finite systems.  Rather than impose
boundary conditions, we stop each simulation sequence whenever the first
particle reaches a boundary.

\begin{figure}
\begin{center}
\includegraphics[width = 8cm]{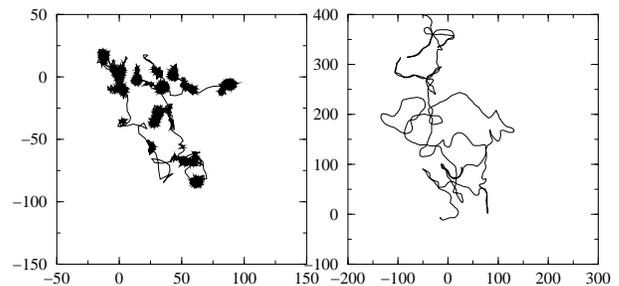}
\end{center}
\caption{
Left: A trajectory for $\gamma=0.1$ over $t=250000$ time units.
Right: A trajectory for $\gamma=0.0001$ over $t=2000$ time
units. The characteristic length scale of the potential is
$\lambda'=4$. Note the different scales in the two panels.
}
\label{fig10}
\end{figure}

Typical trajectories
associated with this potential at high and low friction are shown in
Fig.~\ref{fig10}. Again, the trajectories are entirely different, that
associated with high friction consisting of extremely short steps followed by
random changes in direction while that associated with low friction
again shows a pronounced directional persistence.  
\begin{figure}
\begin{center}
\includegraphics[width = 8cm]{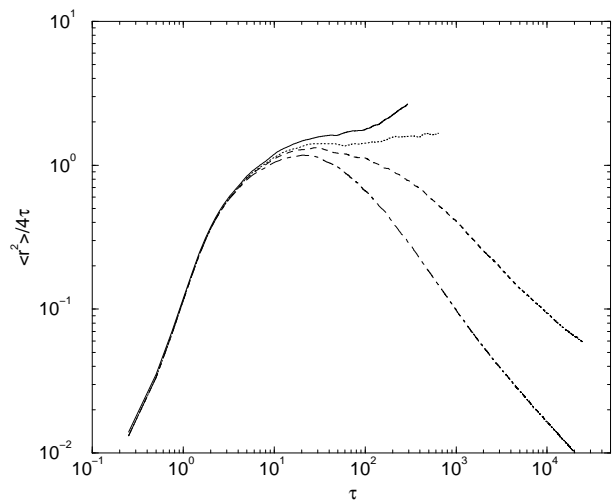}
\end{center}
\caption{
Mean square displacement for a particle in the random potential
with Maxwellian initial velocity distribution, 
for $\gamma=0.0001$ (solid line), $0.001$ (dotted line),
$0.003$ (dashed line) and $0.008$ (dot-dashed line)
}
\label{fig11}
\end{figure}

\begin{figure}
\begin{center}
\includegraphics[width = 8cm]{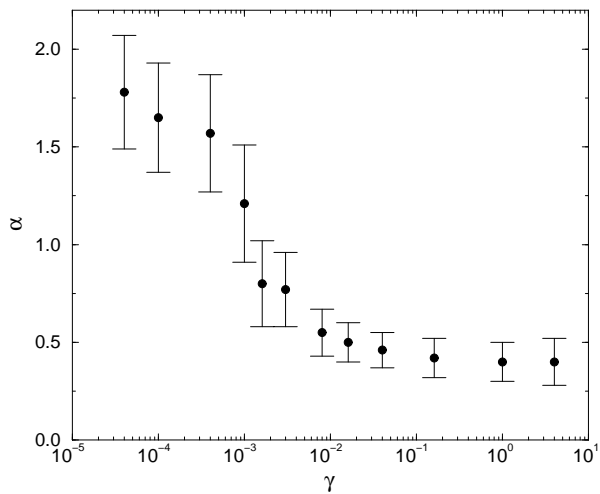}
\end{center}
\caption{
Exponents $\alpha$ versus friction coefficient $\gamma$. Each point is
obtained from an average over $20$ realizations of the random potential
and $500$ particles per realization.  The error bars indicate the
standard deviation over these statistics.
\label{fig12}
 }
\end{figure}

An analysis of the
exponents in the mean square displacement formula shows the
entire range of behaviors from subdiffusive to superdiffusive with
changing friction.
In Fig.\ref{fig11} we show $\langle r^2(\tau)\rangle$, averaged over
$5000$ particles,
as a function of time for several values of $\gamma$.
A detailed analysis of these trajectories presents a rich phenomenology
of possible different regimes.  In the overdamped regime we observe
clearly subdiffusive behavior ($\alpha<1$), already known
from overdamped simulations presented earlier~\cite{romero2}. 
An interesting  outcome is the
superdiffusive ($\alpha>1$) behavior seen for small values of $\gamma$. 
This again is a signature of the L\'evy-walk-type behavior. 
The exponent $\alpha$ as a function of $\gamma$ is plotted
in Fig.~\ref{fig12}.
Each exponent is obtained as an average over $500$ particles in each of
$20$ different realizations of the random potential, the error bars
indicating the standard deviations over these statistics. 
For large $\gamma$
our simulations run for a long time and the value of $\alpha$ has been
calculated on the basis of the behavior of $\langle r^2\rangle$ over the
last time decade of the simulation for each run.  In the medium and
low damping cases
the simulations do not extend over such long times (since each
simulation ends when a particle reaches the system boundary).
The standard
deviations are larger for small $\gamma$.  In part this is due to the
fact that for small $\gamma$ it is more likely for a particle to reach
the boundary of the system during the time of a run.  Not only are the run
times shorter for smaller $\gamma$, but they are also more broadly
distributed as $\gamma$ decreases.  In any case, the trend in the
behavior of $\alpha$ as a function of the friction is clear.
An unexpected result is the absence of a diffusive regime
($\alpha \approx 1$) except in a small range of values. 

\begin{figure}
\begin{center}
\includegraphics[width = 9cm]{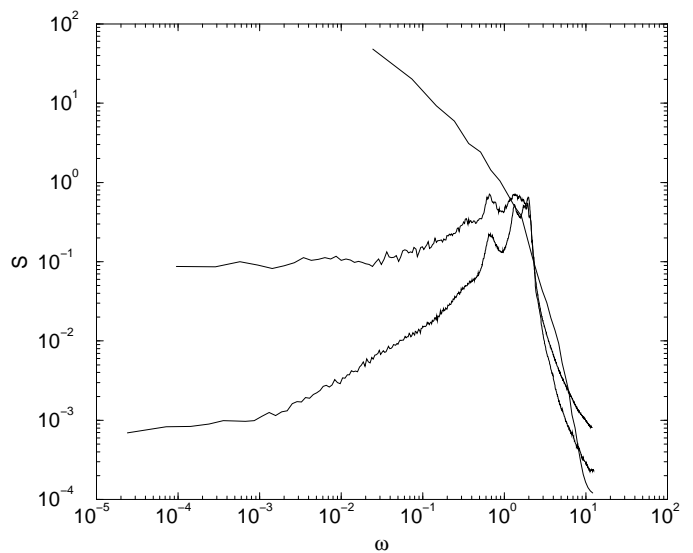}
\end{center}
\caption{$S(\omega)$ for $\gamma=0.0001$, $0.003$ and $0.08$
(decreasing order) in the random potential with Maxwellian
initial velocity distribution.
}
\label{fig13}
\end{figure}

The behavior of $S(\omega)$ for the random potential is seen
in Fig.~\ref{fig13} for several
values of $\gamma$. For small $\gamma$ the spectral
density grows with decreasing frequency, indicating superdiffusion,
while at large $\gamma$ it decays with decreasing $\omega$, which is a 
signature of subdiffusion. For the intermediate value $\gamma = 0.003$
(close to the value at which $\alpha \approx 1$ is observed in
Fig.~\ref{fig12}) it flattens, as appropriate for diffusive behavior
at long times.

\begin{figure}
\begin{center}
\includegraphics[width = 9cm]{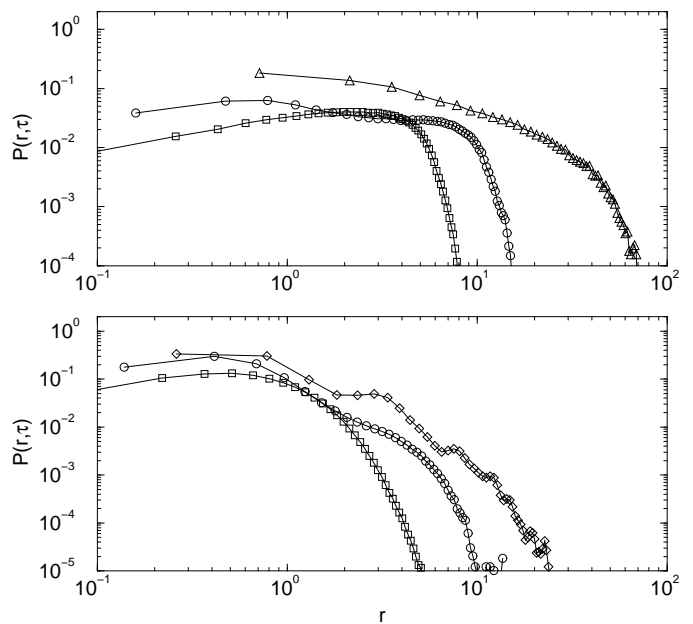}
\end{center}
\caption{
Log-log plot of $P(r,\tau)$ for $\gamma=0.0001$ (top) and for
$\gamma=0.08$ (bottom) and various different values of
time intervals: $\tau= 10$ (squares),
$20$ (open circles), $100$ (triangles), $1000$ (rhombuses).
}
\label{fig14}
\end{figure}

In Fig.~\ref{fig14} the pdf $P(r,\tau)$ for the random potential is plotted. 
This pdf looks very different from the one for a periodic potential.
Nevertheless, for the case of small friction it still exhibits some features
of L\'evi-walk-like dynamics. The main difference here is the absence of the
pronounced central peak due to trapping in the nearest potential well.
This trapping is still evident in the high friction trajectory 
in Fig.~\ref{fig10}. However, for small friction the corresponding peak
in the pdf is smoothed out by the broad distribution of depths
of the sampled minima. Instead, the distribution resembling a smoothed
form of the genuine ballistic L\'evy-walk pdf appears, characterized 
by the central power law dip and side peak. The side peak moves
outward with a constant velocity and becomes broadened by scattering
and trapping.  At long times the pdf develops the 
tent-like central peak typical of trapping.

\section{Conclusions}
\label{sec:conclusions}
We have investigated the behavior of an ensemble of particles
in a two-dimensional
potential subject to thermal fluctuations described by ordinary Langevin
dynamics.  Our purpose has been to determine whether such an ordinary
description is sufficient to produce the entire range of
behaviors, from subdiffusive to superdiffusive, that has been observed
in the motion of molecules on
surfaces~\cite{nano,ala,phd,prl1,phystoday,prl2,jensen,levi,older,henry,condmat,chen}. 
Indeed, we found that it is not
necessary to add any further assumptions to the model to observe the
full range of behaviors, at least on intermediate time scales.  For a
given surface potential one can introduce scaled parameters such that the only
control parameters in a periodic potential are the temperature
(which we have held fixed in this analysis) and the dissipation
parameter $\gamma$. In the case of a random potential we introduced an
additional parameter (which we also hold fixed in this analysis) to
characterize the intensity of the spatial correlation function of the
potential variations.

For the periodic surface potential we found that in the underdamped
regime the motion of the particles includes a ballistic range that
can extend over many decades of time.  The probability distribution
function of the displacement of the particles exhibits a structure
reminiscent of that associated with L\'evy walks.  The pdf has a peak at
short distances that arises from those particles that are trapped in
their original well. The pdf also exhibits a hump that moves outward
linearly with increasing time, and that is due to those particles
whose energy is sufficiently high for ballistic motion over the time
of observation.  As these particles in turn get trapped, the amplitude
and width of this hump decrease.  Particles that start out moving
ballistically eventually lose enough energy through damping to get
trapped, and these progressive trapping events give rise to a
power law intermediate regime connecting the trapped particle peak and
the ballistic particle hump.  These behaviors are not observed in the
case of high dissipation.  The long stretches of essentially ballistic
motion of the more energetic particles also make themselves apparent
in the velocity power spectrum.  We have ascertained that the
L\'evy-like features in the low damping case arise from a subset of
energetic particles by confirming that throughout the evolution the
velocity distribution of particles remains Maxwellian if the initial
distribution is Maxwellian.  If initially the particles have zero
velocity, the distribution rather
quickly settles into an essentially Maxwellian distribution.  Thus,
we stress again that in our model these features arise not because the
average particle is subjected to any esoteric fluctuations, but rather
from the fact that the energetic particles in an ordinary Maxwell
distribution can move over long distances when the system is
underdamped.  At long times (low frequencies) the motion of the
particles eventually becomes diffusive.  This is most clearly seen in
the behavior of the mean square displacement (which eventually grows
linearly with time), and in the plateau of the velocity power spectrum
at low frequencies.  

While most of the results reported above are obtained numerically, we
are able to calculate the diffusion coefficient $D$ that describes the
long-time behavior of the system analytically for high damping (a
well-known result) and also for low damping.  The agreement of our
analytic results for $D$ with numerical ones is good over the entire range
of damping coefficients.

The case of random potential gives rise to a large variety of regimes
ranging from superdiffusion to subdiffusion. While the superdiffusive
behavior is probably a transient, just as in the periodic 
potential, subdiffusion due to multiple trapping is known to be a
true asymptotic behavior when the temperature is low. However, 
in the case of a random potential the restrictions
imposed by our numerical procedure are
tighter, so that the true asymptotic behavior is difficult to reach.  
At short times and for weak damping the L\'evy-walk-like features
arising from long stretches of motion of energetic particles
are still evident in our simulations.
For strong damping this regime is pertinent only for the
shortest times and then
crosses over to that typical of subdiffusive, dispersive transport. 

\section{Acknowledgments}
This work was supported by the MCyT (Spain) under project BFM2003-07850,
by the Engineering Research Program of
the Office of Basic Energy Sciences at the U. S. Department of Energy
under Grant No. DE-FG03-86ER13606, and by a grant from the University of
California Institute for
M\'exico and the United States (UC MEXUS) and the Consejo Nacional de
Ciencia y Tecnolog\'{\i}a de M\'{e}xico (CoNaCyT).
A.H.R. acknowledges support from Millennium Initiative,
Conacyt-Mexico, under Grant W-8001.
I.M.S. acknowledges the hospitality of the University of Barcelona
under the CEPBA grant, as well as partial financial support by
the Fonds der Chemischen Industrie.

\appendix

\section{Diffusion coefficient for high and low friction in a
periodic potential}

We start with the assumption that a particle that
has sufficient energy to move away (``escape'') from a
potential well will in general preferentially
move along directions of lowest potential barriers, that is, along the $r_x$
direction at $r_y=1/2$ (or an odd multiple thereof) or along the $r_y$
direction with $r_x=1/2$ (or an odd multiple thereof).  Choosing the
former, we then have the particle moving along a line in the periodic
potential
\begin{equation}
{\cal V}(r_x)\equiv {\cal V}(r_x,r_y=1/2) = - \sin^2{\pi r_x}.
\label{simplify}
\end{equation}
Only those particles with energy $E>0$ can move. 
As they move they lose energy through dissipation, until they become trapped.
We calculate the diffusion coefficient for the ensemble of particles
according to the formula
\begin{equation}
D = \langle l^2\rangle/(2\tau).
\label{diffusion}
\end{equation}
Here $\langle l^2\rangle$ is the mean square displacement 
from the initial well to the well associated with the next
trapping event, $\tau$ is the mean time for such a journey, and the
numerical factor in the denominator is 2 instead of 4
[cf. Eq.~(\ref{eq4})]
because the motion along each stretch is one-dimensional.
The average is over all the particles in the ensemble,
which includes those that initially have sufficient energy to move
away from a potential well and also those that do not, as dictated by the
Maxwell-Boltzmann distribution. For low-energy particles to be able to
jump out they must acquire enough energy from the thermal fluctuations.

The mean time $\tau$ in general includes the time for a particle to
acquire enough energy ($E>0$) to move out of a well, and the time to
become trapped again in a new well once it has escaped.  
We can calculate $\tau$ using arguments for the calculation of
transition rates  of the Kramers problem ~\cite{Hanggi} by associating
it with the inverse of a transition rate that consists of two
contributions:
\begin{equation}
\tau = \frac{1}{2k} = \frac{1}{2k^{tst} \kappa}.
\end{equation}
One contribution is the transition state theory rate,
\begin{equation}
k^{tst} = \frac{\omega_0}{2\pi} e^{-1/{\cal T}},
\end{equation}
where $\omega_0$ is the angular frequency at the well bottom, which
in our dimensionless variables and for our potential is given by
$\omega_0\equiv [{\cal V}''(1/2)]^{1/2} = \pi\sqrt{2}$. 
The other is the transmission factor $\kappa$, which depends
in a non-trivial way on the damping.

Consider first the case of high friction.
In the overdamped regime (large $\gamma$) we assume that
$\langle l^2\rangle = 1$, and the transmission factor
$\kappa$ has the well-known form~\cite{Hanggi},
\begin{equation}
\kappa = \frac{1}{\omega_b} \left[ \left(
\frac{\gamma^2}{4}+\omega_b^2\right)^{1/2} -\frac{\gamma}{2}\right],
\end{equation}
where $\omega_b^2= {\cal V}''( r_b)= 2 \pi^2$ and $r_b$ is the position of
the maximum of the potential.
Thus the diffusion coefficient in this regime is
\begin{eqnarray}
D &=& \frac{1}{2\pi} \left[ \left(
\frac{\gamma^2}{4}+2\pi^2\right)^{1/2} -\frac{\gamma}{2}\right]
e^{-1/{\cal T}}
\nonumber\\
&\sim& \frac{\pi}{\gamma}e^{-1/{\cal T}}.
\label{diffusionover}
\end{eqnarray}

Now consider the diffusion coefficient in the small-$\gamma$ limit
or infradamped regime.  In this regime a more elaborate analysis
is required to calculate $\langle l^2 \rangle$.
The equations of motion of a particle in the potential
Eq.~(\ref{simplify}) can be expressed in
the standard momentum-position form as the pair,
\begin{eqnarray}
\dot{r_x}&=&p\nonumber\\
\dot{p}&=&-\gamma p-{\cal V}'(r_x) +\zeta(\tau).
\end{eqnarray}
In the small-$\gamma$ limit the energy variation of a particle is slow, and it
is more convenient to rewrite the dynamical equations in terms of the
displacement and the energy $E=p^2/2 + {\cal V}(r_x)$:
\begin{eqnarray}
\label{x}
\dot{r_x}&=& \left\{2\left[E-{\cal V}(r_x)\right]\right\}^{1/2}\\
\label{E}
\dot{E}&=& -2\gamma \left[E-{\cal V}(r_x)\right] +
\left\{2\left[E-{\cal V}(r_x)\right]\right\}^{1/2}\,\zeta(t).
\nonumber\\
&&
\end{eqnarray}
Two quantities needed for our estimate of the diffusion coefficient can be
calculated from these dynamical equations.  One is the time $\tau(E)$
that it takes a particle of energy $E$ to traverse a unit distance
(the scaled spatial period).
The other is the energy $\Delta E$ lost by the particle when it traverses
this distance~\cite{sancho}.  To calculate $\tau(E)$ we assume that the energy
$E$ remains fixed during the traversal, so that we can simply integrate
Eq.~(\ref{x}) for constant $E$.  With the potential~(\ref{simplify})
one obtains a standard integral:
\begin{equation}
\tau(E)=\int_0^1 \frac{dr_x}{\sqrt{2\left[E-{\cal V}(r_x)\right]}} =
\frac{\sqrt{2}\lambda}{\pi\sqrt{1+E}}\mathcal{K}
\left(\sqrt{\frac{1}{1+E}}\right),
\end{equation}
where $\mathcal{K}$ is the complete elliptic integral of the
first kind~\cite{gradshteyn}. 
For $E\ll 1$ the elliptic integral can be approximated by a logarithmic
leading term, which leads to,
\begin{equation}
\tau(E) \approx \frac{1}{\pi\sqrt{2}} \ln \frac{16}{E}.
\end{equation}

Next, we estimate the energy loss that occurs during this traversal.  This
has been calculated in many ways, but a particularly transparent
argument~\cite{sancho} is to neglect the fluctuations in Eq.~(\ref{E})
(since particles above the barrier primarily lose energy)
and integrate over a time interval
$\tau(E)$. The $r_x$ dependence in Eq.~(\ref{E}) even after this approximation
still poses a problem because it leads to incomplete elliptic integrals
amenable only to numerical integration.  Since $r_x$ changes rapidly compared
to $E$, it is reasonable to perform an average of the $r_x$-dependent term
over a traversal from $0$ to $1$.  This leads to the approximate
equation,
\begin{equation}
\dot{E}=-2\gamma F(E)
\end{equation}
where,
\begin{equation}
F(E)\equiv \int_0^1 dr_x P_E(r_x) [E-{\cal V}(r_x)]
\end{equation}
with,
\begin{equation}
P_E(r_x) = \frac{1}{\tau(E)\left[E-{\cal V}(r_x)\right]}.
\end{equation}
The result is,
\begin{equation}
F(E) = \frac{\sqrt{(1+E)}}{\sqrt{2}\pi
\tau(E)}\mathcal{E}\left(\sqrt{\frac{1}{1+E}}\right),
\end{equation}
where $\mathcal{E}$ is the complete elliptic integral of the
second kind~\cite{gradshteyn}. 
Retention of terms to leading order in $E$ and $\Delta E/E$ leads to
\begin{equation}
\Delta E\approx -\frac{{\gamma}\sqrt{8}}{\pi},
\end{equation}
which is independent of $E$.

From the result for $\Delta E$ we can now calculate the mean square
distance $\langle l^2\rangle$ traveled by a particle that starts
with an energy above the barrier.  Since the particle loses energy $\Delta E$
in each traversal of a unit distance, the distance traveled by such
a particle of initial energy $E$ before being trapped is,
\begin{equation}
l(E)  \approx -\frac{E}{\Delta E} \approx \frac{\pi
E}{\gamma\sqrt{8}}.
\end{equation}
Averaging over a Boltzmann distribution of energies, we obtain,
\begin{equation}
\langle l^2 \rangle = (k_BT)^{-1}\int_0^\infty dE\; l^2(E) e^{-E/{\cal T}} =
\frac{(\pi {\cal T})^2}{4\gamma^2}.
\label{lsquared}
\end{equation}

When a
particle of positive energy $E$ moves from one potential well to the next,
it does so in a time $\tau(E)$ and it loses energy $\Delta E$ in the
process.  If its energy is now below the barrier, it will be trapped.
If not, it will keep moving, and will take a time $\tau(E-\Delta E)$ to
get to the next potential well.  The particle moves on until its
energy is too low to continue moving.  The energies have a Boltzmann
distribution, so different particles will take different times to be
trapped both because they are moving progressively more slowly and
because they get trapped at different times in their journey. 
Taking these effects into account, one finally arrives at the
following expression for the transmission factor $\kappa$ \cite{sancho}: 
\begin{equation}
\kappa = \tanh\left(\frac{|\Delta E|}{2 {\cal T}}\right)\approx \frac{-\Delta E}
{2 {\cal T}},
\end{equation}
where the last expression holds for very low damping.  We
thus finally obtain the following result for the
diffusion coefficient in this regime:
\begin{equation}
D \sim \frac{\pi {\cal T}}{4\gamma}e^{-1/{\cal T}}.
\label{diffusioninfra}
\end{equation}

\end{document}